\documentstyle{article}

\def\beq{\begin{eqnarray}}
\def\eeq{\end{eqnarray}}
\def\bea{\begin{eqnarray*}}
\def\eea{\end{eqnarray*}}

\def\centeron#1#2{{\setbox0=\hbox{#1}\setbox1=\hbox{#2}\ifdim
\wd1>\wd0\kern.5\wd1\kern-.5\wd0\fi
\copy0\kern-.5\wd0\kern-.5\wd1\copy1\ifdim\wd0>\wd1
\kern.5\wd0\kern-.5\wd1\fi}}
\def\ltap{\;\centeron{\raise.35ex\hbox{$<$}}{\lower.65ex\hbox{$\sim$}}\;}
\def\gtap{\;\centeron{\raise.35ex\hbox{$>$}}{\lower.65ex\hbox{$\sim$}}\;}

\newcommand{\newc}{\newcommand}
\newc{\qbar}{{\overline q}}
\newc{\Kahler}{K\"ahler }
\newc{\deltaGS}{\delta_{\rm GS}}

\begin{document}
\begin{titlepage}
\begin{flushright}
{\large arXiv:1012.2836 [hep-th] \\
SCIPP 10/07\\
}
\end{flushright}

\vskip 1.2cm

\begin{center}

{\LARGE\bf 
Supersymmetry From the Top Down}

\vskip 1.4cm

{\large Michael Dine}
\\
\vskip 0.4cm
{\it Santa Cruz Institute for Particle Physics and
\\ Department of Physics, University of California,
     Santa Cruz CA 95064  } \\

\begin{abstract}
If supersymmetry turns out to be a symmetry of nature at low energies, the first order of business to measure the soft breaking parameters.  But one will also want to understand
the symmetry, and its breaking, more microscopically.  Two aspects of this problem constitute the focus of these lectures.  First, what sorts of dynamics might account for
supersymmetry breaking, and its manifestation at low energies.  Second, how might these features fit into string theory (or whatever might be the underlying
theory in the ultraviolet).  The last few years have seen a much improved understanding of the first set of questions, and at least a possible pathway to address the second.
\end{abstract}

\end{center}

\end{titlepage}

\tableofcontents \clearpage

\section{String Theory at the Dawn of the LHC Era}

We are on the brink of a new era in particle physics: the LHC program is finally underway.  The missing element of the Standard Model,
the physics responsible for electroweak symmetry breaking, is about to be discovered.  This may be a simple Higgs field, as predicted
in the minimal version of the theory.  But there are good reasons
to think that much more dramatic discoveries may be in store. The hierarchy problem
provides the most compelling
argument that a much richer set of phenomena should be revealed at the
TeV scale.  Theorists have explored a range of possibilities.  Considerations of four dimensional effective field
theory point towards supersymmetry or technicolor, and make a single light Higgs seem highly unlikely.  String theory, understood broadly as
some underlying theory of quantum gravity, may incorporate either or both of these, and points to additional possibilities, such as large extra dimensions and warping.  On the
other hand, string theory, and in particular the
notion of a {\it string theory landscape}\cite{douglas,susskind,anthropichiggs1,anthropichiggs2,anthropichiggs3,anthropichiggs4}, suggests a solution to the hierarchy problem which would lead to a single lonely Higgs.

In these lectures, we will take a ``top down" view of possible physics at the LHC.  We
will focus mainly on supersymmetry, for reasons of time and because this is the case for which we know how to make the most concrete
statements and models.  Supersymmetry has well-known virtues:
\begin{enumerate}
\item  It offers a solution to the hierarchy problem.
\item  It leads to unification of the gauge couplings.
\item  It often provides a suitable candidate for the dark matter.
\item  It is often present in string theory.
\end{enumerate}
Our enthusiasm for supersymmetry, however, should be tempered by the realization that from existing data -- including early LHC data -- there are, as we will
discuss, reasons for skepticism.

Our approach will be ``top down" in the sense that we will ask whether microscopic considerations -- physics at distance scales much shorter than
those which will be probed by the LHC -- point towards some particular set of phenomena at TeV scales.  We will consider two
arenas for this problem:  the dynamics responsible for supersymmetry breaking, and string theory.

\subsection{Two Aspects of the Hierarchy Problem}

Discussions of supersymmetry often begin with the observation
that supersymmetry readily insures the cancelation of the quadratic divergences, which are the most striking manifestation of the
hierarchy problem within the Standard Model (SM).
But at a more primitive level, the hierarchy problem is the
question:  why is the scale of the weak interactions so far below that of gravity
or grand unification?  Supersymmetry can explain\cite{wittendsb}, not only the absence of quadratic divergences, but
the appearance of a small scale.  Essential to this are the so-called non-renormalization theorems, which were described
in Seiberg's lectures (see also \cite{dinebook}).
These  follow from the holomorphy of the gauge couplings and the superpotential.  As a result of these
theorems,  if supersymmetry is
unbroken classically, it is unbroken to all orders of perturbation theory.  These same theorems,
however, indicate that it can sometimes be broken beyond perturbation theory, by effects which
are exponentially small in some weak coupling.  The search for such effects will be a principal theme of these lectures.

\subsection{Reasons for Skepticism}

While supersymmetry has many attractive features, and has established a large and devoted following (it is almost ubiquitous in the
string theory lectures at this school), there are reasons to question
whether it is responsible for the physics of electroweak symmetry breaking, and whether it should make its appearance at the LHC.
Among these are:
\begin{enumerate}
\item  The ``Little hierarchy" problem.  The simplest approach to implementing supersymmetry at low energies,
the so-called ``Minimal supersymmetric standard model" (MSSM),  predicts that there should be a neutral Higgs with mass lighter
than $M_Z$, up to radiative corrections.  These corrections can be substantial, and alternative models can modify the
simplest prediction, but the current limit on the Higgs mass is uncomfortably large.  Many implementations of supersymmetry
require, for example, that the stop quark mass should be greater than $800$ GeV.  Tevatron limits from direct searches place (model-dependent)
limits of several hundred GeV on squarks and gluinos, and the LHC is already setting much more stringent limits (or make a discovery).
\item  Unification:  while unification is successful at the level of effective field theory, it is unclear why it should
be  generic in string theory.  The string constructions described at this school, for example, do not predict unification of couplings,
for typical values of the moduli.  So this more microscopic viewpoint is troubling for one of the seeming successes
of supersymmetry.
\item  There is another hierarchy, for which supersymmetry, or any comparable type of new
physics, fails to offer any solution:  the cosmological constant (c.c.).  Here string theory provides some
guidance, and some cause for concern.  String theory, as a theory of gravity\footnote{Defining precisely what we mean by string theory
is problematic; in these lectures, we will use the term loosely to refer to whatever may be the underlying theory of quantum gravity,
with the understanding that theories of strings provide examples of consistent quantum gravity theories.}, must explain the exceedingly small value of the cosmological constant (dark energy).
The dark energy represents a far more striking failure of dimensional analysis than $M_W/M_p$.
To date, the only plausible explanation is provided by the notion of a {\it landscape}\cite{bankscc,weinbergcc,bp}, the possibility that the theory possesses a vast
array of states, of which states with small c.c. are picked out by anthropic considerations.  While hardly established, this possibility
-- and the problem of the c.c. itself -- raises the worry that there are solutions to problems of hierarchy which cannot be understood
in the framework of low energy effective field theory.  In the case of a landscape, one would expect a distribution of possible values
of the Higgs mass, from which what we (will) observe might be selected by some mechanism\cite{douglas,susskind,anthropichiggs1,anthropichiggs2,anthropichiggs3,anthropichiggs4}.
\end{enumerate}

Despite these cautionary notes, there are several reasons for  (renewed) optimism, which we will touch upon in these lectures:
\begin{enumerate}
\item  The study of metastable susy breaking, initiated by work of Intriligator,
Shih, and Seiberg (ISS)\cite{iss}, has opened rich possibilities for model building with dynamical
breaking of supersymmetry (earlier models were implausibly complex).
\item  Supersymmetry, even in a landscape, can account for hierarchies, as in traditional thinking about naturalness ($e^{-{8 \pi^2 \over g^2}}$)\cite{branches}.
\item  Supersymmetry, in a landscape, accounts for stability -- i.e. the very existence of (metastable) states\cite{dinestability}.
\end{enumerate}

These notes represent the content of three lectures on low energy supersymmetry.  The first two focus on issues in field theory, the third on the question
of supersymmetry in string theory.
\begin{itemize}
\item  Lecture 1:  Low Energy Supersymmetry.  This lecture reviews the basics of supersymmetry and its (metastable) breaking,
the Minimal
Supersymmetric Standard Model (MSSM) and its extensions.
\item  Lecture 2 is devoted to  microscopic models of supersymmetry breaking and its Mediation.
\item  Lecture 3 takes up the question:  ``What might we mean by a string phenomenology?"
 This is a question both broad and difficult to formulate precisely.  We will limit our considerations
 to asking:  does low energy supersymmetry emerge as a prediction of string theory?  While hardly settling the issue,
 we offer arguments, and ask if they lead to more specific predictions about the low energy spectrum.
\end{itemize}

\section{
Lecture 1. Low Scale Supersymmetry and Its Breaking}

In this lecture, we consider:
\begin{enumerate}
\item  Some features of $N=1$ Supersymmetry
\item  Metastable vs. stable supersymmetry breaking in the framework of simple models.
\item  The MSSM.
\item  Gauge Mediated models.
\end{enumerate}

\subsection{A Brief Supersymmetry Review}

It is worth reviewing some
basic  features of $N=1$ theories.  This discussion is necessarily brief; more detail can be found in many texts and
review articles, for example\cite{wb,gates,weinbergbook,tata,drees,aitchison, binetruy,dinebook,terning}.
First, already in global supersymmetry, the supersymmetry algebra already connects an internal,
fermionic, symmetry with space-time.
Denoting the supersymmetry generators, $Q_\alpha$,
they obey the algebra:
\beq
\{Q_{\alpha},Q^*_{\dot
\beta}\}=2\sigma^{\mu}_{\alpha \dot \beta} P_{\mu}
\label{susyalgebra}
\eeq
Tracing in the Dirac indices one has:
\beq
Q_alpha^* Q_\alpha + Q_\alpha Q_\alpha^* = P^0.
\eeq

I will assume familiarity with superspace, and follow
the notation of \cite{wb}.  Without gravity, the
effective theory
should consist of fields with spin at most one.
This permits only two types of supermultiplets (superfields):
\begin{enumerate}
\item
Chiral fields:  these consist of a complex scalar and a Weyl fermion
\beq
\Phi(x,\theta) = \phi(y) +\sqrt{2} \theta \psi(y)+ \theta^2 F,
\eeq
with $y^\mu=x^{\mu} + i
\theta \sigma^{\mu} \bar \theta$.
\item  Vector fields:  These describe fields of spin $1/2$ and spin one.  In superspace,
the field $V$ satisfies the condition $V = V^\dagger$ (it is a {\it real} superfield).   In the case of a $U(1)$ gauge symmetry, the
superspace form of the gauge transformation is:
\beq
V \rightarrow V + \Lambda + \Lambda^\dagger
\eeq
where $\Lambda$ is a chiral field; this has a straightforward generalization
for non-Abelian gauge symmetries.  In a general gauge, $V$ contains a number of unphysical degrees of freedom,
but in the {\it Wess-Zumino} gauge it describes a gauge boson, and a Weyl fermion, and an auxiliary field $D$:
\beq V=  - \theta
\sigma^{\mu} \theta^* A_{\mu} +i \theta^2 \bar \theta \bar
\lambda - i \bar \theta^2 \theta \lambda + {1 \over 2} \theta^2
\bar \theta^2 D .
\eeq
For
our purposes, it is enough to consider the gauge covariant, spin-1/2 chiral field, $W_\alpha$
\beq
W_\alpha = \lambda_\alpha + \theta_\beta \left [\delta^\beta_\alpha D + (\sigma^{\mu \nu})_{\alpha}^{ \beta} F_{\mu \nu}\right ] + \dots
\eeq
\end{enumerate}
$F$ and $D$ are auxiliary fields, i.e. they appear in the lagrangian without derivatives, and so their
equations of motion are simply algebraic; they play an important role as they are order parameters for supersymmetry breaking.
These expressions generalize immediately to non-Abelian theories, thinking of $V$, $F_{\mu \nu}, ~\lambda$, etc. as matrix
valued fields.

The remarkable properties of supersymmetric field theories arise from the highly restricted form of any would-be supersymmetric
lagrangian.
At the level
of terms with two derivatives, ${\cal L}$ is specified by three functions:
\begin{enumerate}
\item  The superpotential, $W(\Phi_i)$, a {\it holomorphic} function of the chiral fields.
\item  The Kahler potential, $K(\Phi_i,\Phi_i^\dagger)$
\item  The gauge coupling functions, $f_a(\Phi_i)$, again holomorphic functions of the fields (one such function for
each gauge group).
\end{enumerate}
The lagrangian density has the form, in superspace:
\beq
\int d^4 \theta K(\Phi,\Phi^\dagger) + \int d^2 \theta \left (f_a(\Phi) W_{\alpha a}^{2} + W(\Phi) \right )
\eeq
We will discuss the component lagrangians further below, but the important point is that they are fully determined
by these functions and their derivatives.

\subsubsection{Renormalizable Interactions}

In thinking about effective theories, either as the low energy limits
of theories which break supersymmetry, or of string theories, we will often be interested in general functions $K$, $W$ and $f$.
But it is is instructive to begin by first restricting our attention to the case of renormalizable theories.
In this case, $K = \sum \Phi_i \Phi_i^\dagger$, $f_a = -{1 \over 4 g_a^2} $ and $W$ is at most cubic.
I will leave the details of the component lagrangian for textbooks and focus here on the scalar potential:
\beq
V = \sum \vert F_i \vert^2 + {1 \over 2} \sum D_a^2,
\eeq
where
\beq
F_i  = {\partial W \over \partial \Phi_i}~~~~~ D^a = \phi_i^* T^a \phi_i.
\eeq
From the basic commutation relations, eqn. \ref{susyalgebra}, we see that supersymmetry is spontaneously broken if and only if the vacuum energy
is non-zero.
Classically, supersymmetry is {\it unbroken} if $\langle F_i \rangle  =\langle  D_a \rangle = 0 ~\forall~ i,a$; conversely, it is broken if not.
If it is broken, there is a Goldstone fermion (``goldstino")\footnote{The existence of the Goldstino follows from features of the supersymmetry
current, just as for Goldstone bosons; see, for example \cite{weinbergbook,dinebook}.},
\beq
G \propto \langle F_i \rangle  \psi_i + \langle D_a \rangle \lambda_a.
\eeq

\subsubsection{R Symmetries}

In supersymmetry, a class of symmetries known as $R$-symmetries play a prominent role.  Such symmetries can be continuous or discrete.  Their
defining property is that they transform the supercurrents,
\beq
Q_\alpha \rightarrow e^{i \alpha} Q_\alpha~~~~Q_\alpha^* \rightarrow e^{-i \alpha} Q_\alpha^*.
\eeq
Necessarily, the superpotential transforms as $W \rightarrow e^{2 i \alpha}$ under this symmetry.

For the question of supersymmetry breaking, the importance of these symmetries is embodied in a theorem of Nelson and Seiberg\cite{nelsonseiberg}:
In order that a generic lagrangian (one with all terms allowed by symmetries) break supersymmetry, the theory
must possess an $R$ symmetry (and in a theory with a spontaneously
broken $R$ symmetry, supersymmetry is necessarily broken).   This theorem is easily proven by examining the equations ${\partial W \over \partial \Phi_i}=0$,
and recalling that they are holomorphic; the proof is reviewed in Seiberg's lectures.  I'll consider, instead, some examples,
illustrating a variety of  $R$ symmetric lagrangians.

In general, $W$ has $R$ charge $2$, if $Q_\alpha$ has charge one.
Consider a theory with fields $X_i$, $i=1,\dots N$ with $R=2$, and $\phi_a$, $a = 1,\dots M$, with $R$ charge $0$.
Then the superpotential has the form:
\beq
W = \sum_{i=1}^N X_i f_i(\phi_a).
\eeq
Suppose, first, that $N=M$.  The equations ${\partial W \over \partial \Phi_i}=0$ are solved if:
\beq
f_i = 0;~~~X_i =0.
\eeq
The first set are $N$ holomorphic equations for $N$ unknowns, and generically have solutions.  Supersymmetry is unbroken; there are a discrete
set of supersymmetric ground states.   Typically there will be
no massless states in these vacua.
The $R$ symmetry is also unbroken, $\langle W \rangle = 0$.

Next suppose that $N < M$.  Then the equations $f_i=0$ contain more unknowns than equations; they generally have an $M-N$  (complex) dimensional space of solutions,
known as a moduli space.  In perturbation theory, as a consequence of non-renormalization theorems, this degeneracy is not lifted.  There are massless
particles associated with these moduli (it costs no energy to change the values of certain fields).

If $N>M$,  the equations $F_i=0$, in general, do not have solutions; supersymmetry is broken.  These are the O'Raifeartaigh models\cite{oraifeartaigh}.
Now the equations ${\partial W \over \partial \phi_i}=0$ do not determine the $X_i$'s, and classically, there are, again, moduli.
Quantum mechanically, however, this degeneracy is lifted.

\subsubsection{Aside 1:  The non-renormalization theorems}

Quite generally, supersymmetric theories have the property that, if supersymmetry is not broken at tree level, then to all orders of perturbation theory, there are no
corrections to the superpotential and to the gauge coupling functions.  These theorems were originally proven by examining detailed properties of Feynman diagrams,
but they can be understood far more simply\cite{seibergnr,seibergnr1}.

To illustrate,
consider a theory with two chiral fields, $\phi$ and $\Phi$, the first light and the second heavy:
\beq
W = {m \over 2} \Phi^2 + {\lambda \over 3} \phi^2 \Phi + {\lambda^\prime \over 3} \phi^3.
\eeq
First set $\lambda =0$.  Then the theory has an $R$ symmetry under which $\Phi$ has unit $R$ charge, and $\phi$ has $R$ charge $2/3$.
The introduction of $\lambda$ breaks the $R$ symmetry, but we can take advantage of the fact that the superpotential is a holomorphic function of
(the complex parameter) $\lambda$, and think of $\lambda$ as itself the expectation value of a (non-dynamical) chiral field.  Then we can assign
$\lambda$ $R$ charge $-1/3$.  Now consider the effective theory at low energies for $\phi$.  Necessarily, any correction to the superpotential
behaves as $\lambda^n \phi^m$, with $-1/3 n + 2/3 m =2$.  So, for example, the $\phi^4/m$ term has coefficient $\lambda^2$, corresponding to
the leading tree diagram. $\phi^6$ would be a one loop diagram, but the only such diagram is not holomorphic in $\lambda$.    Similarly, higher order
polynomials in $\phi$ would necessarily appear at only one order in coupling, and can be shown to be non-holomorphic.

\vskip .2cm
\noindent
{\bf Exercise:}   To this theory, add $\lambda \Phi^3$ and show, by arguments as above, that only tree diagrams contribute to the
low energy superpotential for $\phi$.

\vskip .3cm

\subsection{Metastable Supersymmetry Breaking}

For many years, it was taken for granted that the ultimate goal of supersymmetry model building was to find
theories with stable, dynamical supersymmetry breaking, and that, suitably coupled to the fields of the MSSM,
one would find acceptable low energy soft breaking.  In 2006,
Intriligator, Shih and Seiberg\cite{iss}, demonstrated a surprising result:  in vectorlike, supersymmetric QCD, for a range of colors
and flavors, not only are there supersymmetric vacua, as expected from the Witten index, but there are {\it metastable} states
with broken supersymmetry.  While the particular example is remarkable and surprising, more
generally this work brought the realization that such metastable supersymmetry breaking is a generic phenomenon.  Indeed, this should
have been anticipated from
the Nelson-Seiberg theorem, which asserts that, to be generic, supersymmetry breaking requires a global, continuous $R$ symmetry.
We expect that such symmetries are, at best, accidental low energy consequences of other features of some more microscopic theory.
In such a case, they will be violated by higher dimension operators, and typical theories will exhibit supersymmetry-preserving ground states.

\subsubsection{O'Raifeartaigh Models}

Let's consider the simplest O'Raifeartaigh model in more detail.  This is a model with two fields, $X,~Y$, with $R$ charge $2$, and
a field, $A$, with $R$ charge $0$.  Imposing a $Z_2$ symmetry restricts the model to:
\beq
W = \lambda X (A^2 - f) + m Y A.
\label{simplestor}
\eeq
In this model, SUSY is broken; the equations:
\beq
{\partial W \over \partial X} = {\partial W \over \partial Y} = 0
\eeq
are not compatible.

If  $f > \mu^2$, the vacuum has $\langle A \rangle =0 = \langle Y \rangle$; $X$ undetermined.
It is easy to work out the spectrum.  For $\langle X \rangle =0$, the fermionic components of $A$
combine with those of $Y$ to form a Dirac fermion of mass $m$, while the scalar components of $A$
have mass-squared $m^2 \pm \lambda F_X$ (the scalar components of $Y$ are degenerate with the fermion).
More generally, one should work out the spectrum as a function of $\langle X \rangle$.

\vskip .3cm
\noindent
{\bf Exercise:}
Work out the spectrum as a function of $\langle X \rangle$, first for $X=0$, and then at least for small $X$.

\vskip .3cm

Quantum effects generate a potential for $X$. At one loop, this is known as the Coleman-Weinberg.   As explained below (section \ref{cwpotential}), one
 finds that the minimum of the potential lies at $\langle X \rangle = 0$.  $X$ is lighter than other fields (by
a loop factor).  The scalar components of $X$ are a ``light pseudomodulus."  The spinor is massless; it is the Goldstino of supersymmetry breaking;
\beq
\langle F_X \rangle =f
\eeq
is the decay constant of the Goldstino.

\subsubsection{Aside 2:  The Coleman-Weinberg Potential}
\label{cwpotential}

The
basic idea of Coleman Weinberg calculations for the pseudomoduli
potentials is simple.  First calculate masses of particles as functions of the pseudomodulus.  Then compute the vacuum energy as a function of $\langle X \rangle$.
At lowest order, this receives a contribution of ${1 \over 2} \hbar \omega$ from each bosonic mode, and minus ${1 \over 2} \hbar \omega$ from each fermion
(due to filling the Fermi sea).  As a result:
\beq
V(X)=\sum (-1)^F \int {d^3 k \over (2 \pi)^3}{1 \over 2} \sqrt{k^2 + m_i(X)^2}
\label{vacuumenergy}
\eeq
Term by term, this expression is very divergent in the ultraviolet; expanding the integrand in powers of $k$ yields terms which
are quartically and quadratically divergent.  Introducing a momentum-space cutoff, $\Lambda$, yields:
$$~~~~= \sum(-1)^f \left  (\Lambda^4 + m_i^2 \Lambda^2 + {1 \over (16 \pi^2)} m_i^4 \ln(\Lambda^2/m_i^2) + \dots\right ).$$
The quartically divergent term vanishes because
there are equal numbers of fermions and bosons.  The quadratically divergent term vanishes because of the
tree level sum rule:
\beq
\sum (-1)^F m_i^2 = 0.
\eeq
This sum rule holds in any theory with quadratic Kahler potential. The last, logarithmically divergent
term must be evaluated, when supersymmetry is broken.  The cutoff dependence of this term is associated
with the renormalization of the couplings of the theory (in the O'Raifeartaigh case, the coupling $\lambda$).

One finds in the O'Raifeartaigh model that the potential grows quadratically near $X=0$, and logarithmically
 for large $X$.  As a result, the $R$ symmetry is unbroken.
Shih has shown that this is quite general\cite{shihr}; if all fields have $R$ charge $0$ or $2$, then the $R$ symmetry is unbroken.  Shih constructed models
including fields with other $R$-charges, and showed that in these
$R$ symmetry is typically broken for a range of parameters.  One of the simplest such theories is:
\beq
W = X_2 (\phi_1 \phi_{-1} - \mu^2) + m_1 \phi_1 \phi_1 + m_2 \phi_3 \phi_{-1}.
\label{shihmodel}
\eeq
We will make use of this in model building shortly.

\subsubsection{Continuous Symmetry from a Discrete Symmetry}

The requirement of a continuous $R$ symmetry in order to obtain supersymmetry breaking is, at first sight, disturbing.  It is generally
believed that a consistent theory of quantum gravity cannot exhibit global continuous symmetries (for a recent discussion of this issue,
see \cite{banksseiberg}).  Discrete symmetries, however, are different; these can be gauge symmetries (for example, they can be
discrete subgroups of a broken continuous gauge symmetry, or discrete remnants of higher dimensional space-times symmetries).  Such
exact symmetries have the potential to give rise to {\it approximate} continuous global symmetries.

As an
example, the continuous symmetry of the OR model might arise as an accidental consequence of a discrete, $Z_N$ {\cal R} symmetry.
This could simply be a subgroup of the $R$ symmetry of the ``naive" model.  For example:
\beq
X \rightarrow e^{4 \pi i \over N} X;~Y \rightarrow e^{4 \pi i \over N} Y
\label{ror}
\eeq
corresponding to $\alpha = {2 \pi \over N}$ in eqn. \ref{ror} above.

For general $N$, this symmetry is enough to ensure that, keeping only renormalizable terms,
the lagrangian is that of equation \ref{simplestor}.  But higher dimension terms can break
the continuous $R$ symmetry.
Suppose, for example, $N=5$.  The discrete symmetry now allows couplings such as
\beq
\delta {W} = {1 \over M^3} \left ( a X^6 + b Y^6 + c X^4 Y^2 + d X^2 Y^4 + \dots \right ).
\eeq
Note that $W$ transforms, as it must, under the discrete $R$
symmetry,  $W \rightarrow e^{4 \pi i \over N} W$.

The theory now has $N$ supersymmetric minima, with
\beq X \sim \left ( \mu^2 M^3 \right )^{1/5} \alpha^k
\eeq
where $\alpha = e^{2 \pi i \over 5}$, $k = 1,\dots,5$.
Classically, the original point near the origin is no longer stationary.  

For large $M$, these vacua are ``far away"
from the origin.  Near the origin, the
higher dimension (irrelevant) operator has negligible effect, so the Coleman-Weinberg calculation, even though
suppressed by a loop factor, gives the dominant contribution to the potential.  The potential still exhibits a local
minimum, however its global minima are the supersymmetric ones.

\subsubsection{Metastability}

The broken supersymmetry state near the origin, at least in the limit of global supersymmetry, will eventually
decay to one of the supersymmetric minima far away.
We can ask how quickly this decay occurs.  We would
need a separate lecture to discuss tunneling in quantum field theory (some remarks
on this subject appear in Banks' lectures in this volume).  Suffice it to say that in models such as those
introduced above, the metastable supersymmetric state can be extremely long lived.
In particular, the system has to tunnel a ``long way" (compared with characteristic energy
scales) to reach the ``true" vacuum.  Thinking (correctly) by analogy to WKB,
the amplitude is exponentially suppressed by a power of the ratio of these scales.  An elementary discussion
appears in \cite{dinemasonreview}.

\subsection{Macroscopic Supersymmetry:  The MSSM and Soft Supersymmetry Breaking}

If one simply writes a supersymmetric version of the Standard Model, it is not hard to show
that supersymmetry cannot be spontaneously broken in a realistic fashion.  So it is generally
assumed that the dynamics responsible for supersymmetry breaking operates at a scale
well above the weak scale, and in particular above the mass scale of the superpartners of ordinary
fields.  At lower energies, one has a supersymmetric theory, consisting of the SM fields and their superpartners,
and perhaps some limited number of additional fields, described by a supersymmetric effective
field theory with explicit soft breaking of supersymmetry.  As a result, we can divide our considerations into ``macroscopic" supersymmetry -- the
phenomenological description of this effective theory -- and microscopic supersymmetry, the detailed
mechanism by which supersymmetry is broken and this breaking is communicated to the partners of the
SM fields.  This section is devoted to this macroscopic picture; then, in section \ref{microphysics} we will turn to the more microscopic questions.

The MSSM is a supersymmetric generalization of the Standard Model(SM).
Its field content and lagrangian are characterized by:
\begin{enumerate}
\item  Gauge group $SU(3) \times SU(2) \times U(1)$; correspondingly there are (twelve vector multiplets.
\item  Chiral field for each fermion of the SM:  $Q_f,\bar U_f, \bar D_f, L_f, \bar E_f$.
\item  Two Higgs doublets, $H_U, H_D$.
\item  The superpotential of the MSSM contains a generalization of the Standard Model Yukawa couplings:
\beq
W_{y} = y_U H_U Q \bar U + y_D H_D Q \bar D + y_L H_D \bar E .
\eeq
$y_U$ and $y_D$ are $3 \times 3$ matrices in the space of flavors.
\end{enumerate}

\subsubsection{Soft Breaking Parameters}

Needless to say, it is important that supersymmetry be broken.  For this purpose, one can try to construct a complete model of spontaneous supersymmetry
breaking, or one can settle for an effective theory which is supersymmetric up to  {\it explicit soft breakings}.   The term ``soft" refers to the fact that these breakings
only have mild effects at short distances; in particular, they do not appreciably affect the renormalizable (marginal) operators, while they are themselves at most
corrected logarithmically.  It is easy to list the possible soft terms\cite{softterms}:
\begin{enumerate}
\item  Mass terms for squarks, sleptons, and Higgs fields:
\beq
{\cal L}_{scalars}= Q^* m_Q^2 Q + \bar U^* m_U^2 \bar U + \bar D^* m_D^2 \bar D
\eeq
$$~~~~+ L^* m_L^2 L
 + \bar E^* m_E \bar E $$
 $$~~~~
+ m_{H_U}^2 \vert H_U\vert^2 + m_{H_U}^2 \vert H_U\vert^2 + B_\mu H_U H_D + {\rm c.c.}
$$
$m_Q^2$, $m_U^2$, etc., are hermitian matrices in the space of flavors.  Each has $9$ real parameters.
\item  Cubic couplings of the scalars:
\beq
{\cal L_A} = H_U Q ~A_U ~\bar U +H_D Q ~A_D ~\bar D
\eeq
$$~~~~ +H_D L ~A_E
 \bar E + {\rm c.c.}
$$
The matrices $A_U$, $A_D$, $A_E$ are complex matrices, each with $18$ real entries.
\item  Mass terms for the U(1) ($b$), $SU(2)$ ($w$), and $SU(3)$ ($\lambda$) gauginos:
\beq
m_1 b b + m_2 w w + m_3 \lambda \lambda
\eeq
These represent $6$ additional parameters.
\item  $\mu$ term for the Higgs field,
\beq
W_\mu = \mu H_U H_D
\eeq
representing two additional parameters.
\end{enumerate}

So we have the following counting of parameters beyond those of the SM:
\begin{enumerate}
\item
$\phi \phi^*$ mass matrices are $3\times 3$ Hermitian ($45$ parameters)
\item
Cubic terms are described by $3 \time 3$ complex matrices ($54$ parameters)
\item
The soft Higgs mass terms add an additional $4$ parameters.
\item
The $\mu$ term adds two.
\item
The gaugino masses add $6$.
\end{enumerate}
So there appear to be $111$ new parameters.

But the Higgs sector of the SM has two parameters.
In addition, the supersymmetric part of the MSSM
lagrangian has symmetries which are broken by the general soft breaking terms (including $\mu$ among the
soft breakings):
\begin{enumerate}
\item  Two of three separate lepton numbers
\item  A ``Peccei-Quinn" symmetry, under which $H_U$ and $H_D$ rotate by the same phase, and the quarks and leptons transform suitably.
\item  A continuous ``$R$" symmetry, which we will explain in more detail below.
\end{enumerate}
Redefining fields using these four transformations reduces the number of parameters to $105$.

If supersymmetry is discovered, determining these parameters,
and hopefully understanding them more microscopically, will be the main business of particle physics for some time.  The phenomenology
of these parameters has been the subject of extensive study; we will focus here on a limited set of issues.

\subsubsection{Constraints on the Soft Breaking Parameters}

Over the years, there have been extensive
direct searches (LEP, Fermilab) for superpartners of ordinary particles,
and these severely constrain the spectrum.  For example, squark and gluino masses must be greater than
$100$'s of GeV, while chargino masses
of order $100$ GeV; early LHC running has already substantially strengthened the gluino limit.
But beyond these direct searches, the
spectrum must have special features to explain
\begin{enumerate}
\item  absence of Flavor Changing Neutral Currents (suppression of $K \leftrightarrow \bar K$, $D\leftrightarrow \bar D$ mixing; $B \rightarrow s + \gamma$,
$\mu \rightarrow e + \gamma$, \dots)\cite{flavorreviews}
\item  suppression of $CP$ violation ($d_n$; phases in $K\bar K$ mixing).
\end{enumerate}
Both would be accounted for if the spectrum is highly degenerate, and CP violating phases in the soft breaking lagrangian are suppressed.
This happens in many gauge mediated models, as we will discuss shortly, and in special regions of some superstring moduli spaces\cite{stringflavor}.
Other possible explanations include flavor symmetries\cite{flavorsusy}.

\subsubsection{The little hierarchy:  perhaps the greatest challenge for Supersymmetry}

While with low energy supersymmetry, radiative corrections to the Higgs mass are {\it far} smaller than within the Standard Model,
current experimental constraints still render them uncomfortably large.
The largest contribution to the Higgs mass arises from top quark loops.  There are two graphs, one with an intermediate
top squark, one with a top quark; they cancel if supersymmetry is unbroken.  The result of a simple computation is
\beq
\delta m_{H_U}^2 = -6 {y_t^2 \over 16 \pi^2} \tilde m_t^2 \ln(\Lambda^2/\tilde m_t^2)
\label{higgsmassshift}
\eeq
Even for modest values of the coupling, given the limits on squark masses, this can be substantial.
The fine tuning seems to be order $1\%$.

\vskip .3cm
\noindent
{\bf Exercise}  Derive eqn. \ref{higgsmassshift}.

\vskip .3cm

But the experimental limit, $m_H > 114$ GeV, poses another problem.  At tree
level, in the MSSM, $m_H \le m_Z$.  This traces to the fact that the quartic
couplings of the Higgs, in the MSSM, are determined by the gauge couplings.  Fortunately
(for the viability of the model) loop corrections involving the top quark can substantially correct the Higgs quartic
coupling, and increase the Higgs mass mass\cite{haberhiggs,ellishiggs,yanagidahiggs}.
The leading contribution is proportional to $\log(\tilde m_t/m_t)$, and is readily calculated:
\beq
\delta \lambda \approx 3{y_t^4 \over 16 \pi^2} \log(\tilde m_t^2/m_t^2).
\label{higgsquarticshift}
\eeq
This is to be compared with the tree level term, of order ${g^2 + g^{\prime ~2} \over 8}$, which is
not terribly large.  Still, evading the LEP bound
typically requires $\tilde m_t > 800$ GeV.  This exacerbates the problem of tuning, which now appears,
over much of the parameter space, to be worse than $1 ~\%$.

\vskip .3cm
\noindent
{\bf Exercise}  Derive eqn. \ref{higgsquarticshift}.

\vskip .3cm

A variety of solutions have been proposed to this problem, and there is not space to review them all here.  I will mention one, which will
be tied to ideas we will develop subsequently.
Suppose that there is some additional physics at a scale somewhat above the scale of the various superparticles.  Then the Higgs coupling
 can be corrected by dimension five term in the superpotential or dimension six in the Kahler potential\cite{bmssm}
\beq
\delta W = {1 \over M} H_U H_D H_U H_D~~~\delta K = Z^\dagger Z H_U^\dagger H_U H_U^\dagger H_U.
\eeq
For plausible values of $M$, and including radiative corrections as well, these couplings can lift the Higgs mass somewhat above the LEP bound.

A possible origin for this operator might be an extra, massive singlet, coupled to the Higgs:
\beq
W_S = {M \over 2}S^2 + \lambda S H_U H_D .
\label{NMSSM}
\eeq
Models with an additional singlet beyond the MSSM fields are known collectively as the ``Next to Minimal Supersymmetric Standard Model", NMSSM.  Model builders
make different assumptions about this theory; most forbid the mass term of eqn. \ref{NMSSM} as unnatural.  We discuss this and related issues
later.

\section{Microscopic Supersymmetry}
\label{microphysics}

Having explored the MSSM and its generalizations, we turn now to more microscopic considerations.  First, we will simply assume that some dynamics is responsible
for supersymmetry breaking, and ask how this breaking might be communicated.  There are many approaches which have been considered,
but we will
focus on two, which have captured the most attention:  Gravity mediation and gauge mediation.

\subsection{Supergravity}

In both cases, we need to know something about supergravity.
We expect supersymmetry to be a local symmetry.  Supergravity
is a non-renormalizable theory; it is necessarily applicable only
over a limited range of energies, and cannot be used for computation
of quantum effects; some ``ultraviolet completion" is required.
But, as we will see, if supersymmetry is broken at a scale well below
$M_p$,, Planck scale
effects can potentially control important aspects of low energy physics, and
these can be described in terms of a lagrangian
with local supersymmetry.   The most general supergravity lagrangian with terms up to two derivatives
appears in\cite{wb,ferraraetal}; a good introduction is also provided
by \cite{weinbergbook}.  Much like the global case, the general
lagrangian is specified by a Kahler potential, superpotential, and gauge coupling functions,
Here, we will content ourselves with describing
some features which will be important for model building and certain more general theoretical issues.

Perhaps most important for us will be the form of the scalar potential:
In units with $M_p = 1$ (here $M_p$ is the reduced Planck mass, approximately $2 \times 10^{18}$ GeV):
\beq
V = e^{K} \left [ D_i W
g^{i \bar i} D_{\bar i} W^*- 3 \vert W \vert^2 \right ].
\eeq
$D_i W\equiv F_i$ is the
order parameter for susy breaking:
\beq
D_i W = {\partial W \over \partial \phi_i} + {\partial K \over \partial \phi_i} W.
\eeq

If supersymmetry is {\it unbroken}, space time is Minkowski (if $W=0$), It is AdS if ($W \ne 0$).
If supersymmetry is broken and space is approximately flat space ($\langle V \rangle = 0$), then
\beq
m_{3/2} =\approx \langle e^{K/2} W \rangle.
\eeq

\subsection{Mediating Supersymmetry Breaking}

The classes of models called ``gauge mediated" and ``gravity mediated" are distinguished principally by the scale at which supersymmetry is broken.
If the $F_i$'s are large enough,  terms in the supergravity lagrangian (more generally, higher dimension operators) suppressed by $M_p$ are important at the weak (TeV) scale.
This requires:
\beq
F_i = D_i W \approx (TeV) M_p \equiv M_{int}^2.
\eeq
For such $F_i$,
we will speak of the supersymmetry breaking as
``gravity mediated".  We will refer to $M_{int} \approx 10^{11}$ GeV as the ``intermediate scale", as it is the
geometric mean of the scale of weak interactions and the Planck scale.  If the scale is lower, we will use the term ``gauge mediated".
 More precisely, gauge mediated
models are models where supersymmetry breaking is transmitted principally through gauge interactions.
In practice, as we will explain shortly, it is difficult to construct low scale models which are {\it not} gauge mediated; this is the rationale
for our terminology.

\subsection{Intermediate Scale Supersymmetry Breaking (``Gravity Mediation")}

In the case of intermediate scale breaking, non-renormalizable couplings
are responsible for the essential features of the physics at low energies.  Such couplings
are inherently sensitive to high scale physics.  Lacking an ultraviolet completion of the theory,
such as a (fully understood) string theory, one can only speculate about the origin and nature of these couplings;
in general, they must be viewed as free parameters.  As a result, there is enormous freedom in building models; one can readily
fill out the full set of parameters of the MSSM.  As a result, one must make strong assumptions about the microscopic physics in order to be consistent with
existing low energy constraints.   If, for example, we have a field, $Z$, responsible
for supersymmetry breaking,
\beq
e^{K/2}F_Z  = m_{3/2}M_p
\eeq
then if $K$ is a polynomial in $Z$ and the other fields in the theory, all terms up to at least those quartic in fields are important in determining the low
energy features of the theory.
Suppose, for example, an O'Raifeartaigh-like model breaks supersymmetry.  Choosing the constant in the
superpotential, $W_0$ is chosen so that the cosmological constant is very small.
\beq
F_Z \equiv D_Z W = {\partial W \over \partial Z} + {\partial K \over \partial Z} W \ne 0
\eeq
along with
\beq
W_0 = {1 \over \sqrt{3}} \langle F \rangle
\eeq
leads to soft masses for squarks, sleptons.  In particular, for the MSSM fields, $\phi_i$, the terms
in the potential:
\beq
V(\phi) \approx  {\partial K \over \partial \phi_i}{\partial K \over \partial \phi_{\bar j}}^* g^{i \bar j} \vert W \vert^2
\eeq
contribute to the mass-squared of all fields an amount of order $m_{3/2}^2 = e^K \vert W \vert^2$.
Couplings such as
 $\int d^2 \theta Z W_\alpha^2$ c an give mass to gauginos.

If the Kahler potential terms for the $\phi_i$ fields are simply
\beq
K_\phi = \sum \phi_i^\dagger \phi_i
\eeq
then all of the $\phi_i$ fields acquire a mass-squared equal to $m_{3/2}^2$.  However, terms in the Kahler potential:
\beq
\delta K = {\gamma^{\bar i j} \over M^2} Z^\dagger Z \phi_{\bar i}^\dagger \phi_j
\eeq
yields flavor-dependent masses for squarks and sleptons.   No symmetry forbids such terms (approximate flavor symmetries might constrain them, however).

\vskip .2cm
\noindent{\bf Exercise:}  Verify that by choice of the $\gamma_{ij}$'s one can explore the full parameter space of the MSSM.

\vskip .2cm.

Other potential difficulties with intermediate scale models include cosmological problems, such as the gravitino overproduction
and moduli problems\cite{moduliproblem,moduliproblem1}.
We will not elaborate these here; cosmology also constrains gauge mediated models.

\subsubsection{Low Scale Supersymmetry Breaking:  Gauge Mediation}

In the low scale case, the soft breaking effects at low energies should be calculable, without requiring further ultraviolet completion; this is the
arena for field theory ``model building."  It is not hard to show that within the MSSM, there is no mechanism which can break
supersymmetry suitably.  So additional degrees of freedom are certainly required.   One can contemplate many possibilities, both for the number
and gauge transformation properties of the fields, and their couplings to MSSM fields.  One faces several challenges:
\begin{enumerate}
\item  Obtaining positive mass-squared for partners of squarks and sleptons.  This turns out to
be achieved simply if the gauge couplings of the MSSM (i.e. the supersymmetric version of the Standard Model gauge
coupling) mediate the breaking of supersymmetry.  Yukawa couplings to new fields associated with supersymmetry breaking
tend to be problematic.
\item  Suppressing flavor changing neutral currents.  This tends to require some sort of flavor symmetry.  If gauge interactions are the
dominant source of squark and slepton masses, one immediately has an approximate flavor symmetry.  With Yukawa couplings to new fields,
the challenges are more serious.
\item  Other model building issues include:  obtaining suitable gaugino masses and a $\mu$ term for Higgs fields.
\end{enumerate}

In the rest of this lecture, we will focus exclusively on gauge mediation\cite{gaugemediationreview}.  First we describe the simplest model of gauge mediation, ``Minimal Gauge Mediation",
which
is remarkably predictive\cite{mgm,mgm1,mgm2,mgm3}.  Then we turn to the general case\cite{ggm}.

\subsubsection{Minimal Gauge Mediation}

The main premiss underlying gauge mediation can be simply described:  in the limit that the gauge couplings vanish, the hidden and visible sectors decouple.\footnote{This
definition was most clearly stated in \cite{ggm}, but some care is required, since, as we will see, additional features are needed for a realistic model.}
Perhaps the simplest model of gauge mediation, known as Minimal Gauge Mediation, involves a chiral field, $X$, whose vacuum expectation value
is assumed to take the form:
\beq
\langle X \rangle = x + \theta^2 F.
 \eeq
 $X$ is coupled to a vector-like set of fields, transforming as $5$ and $\bar 5$ of $SU(5)$:
 \beq
 W = X(\lambda_\ell \bar \ell \ell + \lambda_q \bar q q).
\label{wmgm}
\eeq
For $F<X$, $\ell, \bar \ell, q, \bar q$ are massive, with supersymmetry breaking splittings of order $F$.
The fermion masses are given by:
\beq
m_q = \lambda_q x~~~ m_\ell = \ \lambda_\ell  x
\eeq
while the scalar splittings are
\beq
\Delta m_q^2 = \lambda_q F ~~~~~ \Delta m_\ell^2 = \lambda_\ell F.
\eeq

 In such a model, masses for gauginos are generated at one loop; for scalars at two loops.  The gaugino mass computation
 is quite simple.  The two loop scalar
 masses are not very difficult, as one is working at zero momentum.  The latter calculation
can be done quite efficiently using supergraph
techniques; an elegant alternative uses background field arguments\cite{backgroundmasses,backgroundmasses1}.
The result for the gaugino masses is:
\beq
m_{\lambda_i} = {\alpha_i \over \pi} \Lambda,
\label{gauginomasses}
\eeq

For the squark and slepton masses, one finds
\beq
\widetilde m^2 ={2 \Lambda^2}
[
C_3\left({\alpha_3 \over 4 \pi}\right)^2
+C_2\left({\alpha_2\over 4 \pi}\right)^2
\label{scalarsmgm}
\eeq
$$~~~
+{5 \over 3}{\left(Y\over2\right)^2}
\left({\alpha_1\over 4 \pi}\right)^2 ],
$$
where $\Lambda = F_x/x$.
$C_3 = 4/3$ for color triplets and zero for singlets,
$C_2= 3/4$ for
weak doublets and zero for singlets.

\vskip .2cm
\noindent
{\bf Exercise:}   Derive eqn. \ref{gauginomasses}.

Examining eqns. \ref{gauginomasses}, \ref{scalarsmgm} one can infer the following remarkable features of MGM:
\begin{enumerate}
\item  One parameter describes the masses of the three gauginos and the squarks and sleptons
\item  Flavor-changing neutral currents are automatically suppressed; each of the matrices $m_Q^2$, etc., is automatically proportional to the
unit matrix.  The corrections are tiny, and the $A$ terms are highly suppressed (they receive no one contributions before three loop order).
\item  CP conservation is automatic
\item  This model cannot generate a $\mu$ term; the term is protected by symmetries.  Some further structure is necessary.
\end{enumerate}

\subsubsection{General Gauge Mediation}

Much work has been devoted to understanding the properties of this simple model, but it is natural to ask:
just how general are these features?  It turns out that they are peculiar to our assumption of a single set of messengers
and just one singlet responsible for supersymmetry breaking and R symmetry breaking.
Meade, Seiberg and Shih have formulated the problem of gauge mediation in a general way,
and dubbed this formulation {\it General Gauge Mediation}  (GGM).    They study the problem
in terms of correlation functions of (gauge) supercurrents.  Analyzing the restrictions imposed by Lorentz invariance and supersymmetry
on these correlation functions, they find that the general gauge-mediated spectrum is described by three complex parameters and three real
parameters.  The spectrum can be significantly different than that of the MGM, but the masses are still only functions
of gauge quantum numbers and flavor problems are still mitigated.

The basic structure of the spectrum is readily described.   In the formulas for fermion masses we introduce a separate complex parameter $m_i,~i=1,\dots 3$
for each Majorana gaugino.  Similarly, for the scalars, we introduce a real parameter $\Lambda_c^2$ for the contributions from $SU(3)$ gauge fields, $\Lambda_w^2$
for those from $SU(2)$ gauge fields, and $\Lambda_Y^2$ for those from hypercharge gauge fields:
\beq
\widetilde m^2 = 2
\left[
C_3\left({\alpha_3 \over 4 \pi}\right)^2 \Lambda_c^2
+C_2\left({\alpha_2\over 4 \pi}\right)^2 \Lambda_w^2
+{5 \over 3}{\left(Y\over2\right)^2}
\left({\alpha_1\over 4 \pi}\right)^2 \Lambda_Y^2\right].
\eeq
One can construct models which exhibit the full set of parameters\cite{cdf,seibergallparameters}.  In MGM, the messengers of each set of quantum
numbers each have a supersymmetric contribution to their masses, $\lambda M$, while the supersymmetry breaking
contribution to the scalar masses goes as $\lambda M^2$, so in the ratio the coupling cancels out.  In GGM model building,
additional fields and couplings lead to more complicated relations.

One feature which of MGM which is not immediately inherited by GGM is the suppression of new sources of CP violation.
Because the gaugino masses are independent parameters, in particular, they introduce additional phases which are inherently
CP violating.  Providing a natural explanation of the suppression of these phases is one of the main challenges of GGM model building.

\section{Lecture 2:  Microscopic Models of Supersymmetry Breaking}

In this second lecture, we will continue our consideration of more microscopic models of supersymmetry and its breaking.  This lecture covers:
\begin{enumerate}
\item  Low Energy, Dynamical Supersymmetry Breaking:  A connection to the Cosmological Constant
\item  The importance of Discrete R Symmetries
\item  Gaugino condensation and its generalizations
\item  Building models of Low Energy Dynamical Supersymmetry Breaking
\item  Assessment
\item  A theorem about the superpotential
\end{enumerate}

\subsection{Low Energy Supersymmetry Breaking and the Cosmological Constant}

In this lecture, we will focus on low energy supersymmetry breaking.  While
 we won't consider string constructions per se, we will consider an important connection with gravity:  the cosmological constant.  We will not be attempting to provide
a new explanation, but rather simply asking about the features of the low energy
lagrangian in a world with approximate SUSY and small $\Lambda$.  We will argue that this may be a guide to the microscopic mechanism
of supersymmetry breaking.

With supersymmetry, there is an inevitable connection of {\it low} energy physics and gravity:
\beq
\langle \vert W \vert^2\rangle  =3\langle  \vert F \vert^2 \rangle M_p^2+ {\rm ~tiny}.
\eeq
So not only do we require that $F$ be small, but also $W$.  Why?  A few possible answers have been offered:
\begin{enumerate}
\item  Some sort of accident?  For example, in the KKLT scenario\cite{kklt}, one assumes tuning of $W$ relative to $F$ (presumably anthropically).
\item  R symmetries can account for small $W$  (Banks).  We we will see,  $<W>$ can be correlated naturally
with the scale of supersymmetry breaking.
\end{enumerate}

These remarks
suggest a possible role for $R$ symmetries.   In string theory (gravity theory) such symmetries are necessarily  discrete. and they are,
at least at the level of textbook models, ubiquitous.  They can arise, for example, as discrete subgroups of a higher dimensional
Lorentz group, preserved by compactification.  As such, they are necessarily discrete gauge symmetries, expected to survive
in the quantum theory.
Discrete $R$ symmetries are interesting from several points of view:
\begin{enumerate}
\item  They can account for the small $W_0$ needed to understand the cosmological constant
\item  They can give rise to approximate continuous R symmetries at low energies  which can account for
supersymmetry breaking (Nelson-Seiberg).
\item   They can account for small, dimensionful parameters.
\item They can provide needed suppression of proton decay and other rare processes.
\end{enumerate}

\subsubsection{Continuous R Symmetries from Discrete Symmetries}

Recall that the basic OR model possesses a continuous $R$ symmetry::
\beq
W = X_2(A_0^2 - f ) + mA_0Y_2
\eeq
(subscripts denote $R$ charges).
If, e.g., $\vert m^2 \vert > \vert f \vert$, $F_X = f.$
We have seen that this model can arise
as the low energy limit of a model with a discrete $R$ symmetry:
\beq
X_2 \rightarrow e^{2 \pi i \over N}X_2;~Y_2 \rightarrow e^{2 \pi i \over N}Y_2;~A_0 \rightarrow A_0.
\eeq

This symmetry allows higher dimension (non-renormalizable) terms such as
\beq
\delta W = {X^{N-n} Y^{n+1} \over M_p^{N-2}}.
\eeq
The model has
$N$ supersymmetric vacua, far away from the supersymmetry-breaking vacuum near the origin..
Physics in this vacuum exhibits an approximate, accidental R symmetry.
The state with broken supersymmetry is {\it highly} metastable.

One can treat Shih's model, eqn. \ref{shihmodel}, in a similar fashion.  Coupling the field of that model to
messengers, as in the MGM  (eqn. \ref{wmgm}), one can build a realistic model of gauge mediation.
Additional tunneling instabilities arise as there are now additional supersymmetric vacua, in which
some of the messenger fields are non-vanishing.  Again, however, the desired metastable state
can readily be {\it highly} metastable.  We will discuss these models further later.

\subsection{Retrofitting the O'Raifeartaigh Models}

Up to this point, in these lectures, we have distinguished a notion of ``macroscopic physics", phenomena occurring at the TeV scale,
and ``microscopic physics", associated with supersymmetry breaking.  The notion of ``microscopic", however, already requires some
refinement in light of our discussion of gauge mediation.  Here we have a mass scale associated with the messengers, and a potentially
very different scale associated with the fields which break supersymmetry (the fields $X$ of the various O'Raifeartaigh models, for example).
Even these scales may arise from dynamics at still higher scales.  While such a structure may seem arcane, we will see
that it can be quite natural, and even (automatically) compatible with the order of magnitude of $W$ required to obtain small cosmological constant.

In this section, we will describe a simple strategy for building models with metastable, dynamical supersymmetry breaking,
known as ``retrofitting"\cite{retrofitting}.
This breaking will be induced by dynamics at a higher scale which dynamically breaks a discrete $R$ symmetry, without breaking
supersymmetry.  The prototype for such theories are pure gauge theories, in which gaugino condensation
breaks a $Z_N$ symmetry, in the case of $SU(N)$.  In the following subsection, we first generalize gaugino condensation
to theories which include order parameters of dimension one\cite{dk,kehayias}.  We will then be in a position to build models
in which all dimensionful parameters arise through dimensional transmutation, including the $\mu$ term of the MSSM,
and possible parameters of the NMSSM.  We will see that the expectation value of the superpotential plays an important role.
Since
$W$ transforms under any $R$ symmetry, $\langle W \rangle$ itself is an order parameter for $R$ breaking.  In the context of supergravity
theories, this is particularly important.  We will see that the relations among scales implied by the retrofitting hypothesis are of the correct
order of magnitude to account for the smallness of the cosmological constant; this is not true of many other schemes for supersymmetry
breaking, where additional scales must be introduced by hand.

\subsubsection{Generalizing Gaugino Condensation}

In this section I will assume some familiarity with basic aspects of supersymmetric
dynamics.  One can skip this section, and simply accept the basic result, that one can construct
models in which $R$ symmetry is dynamically broken, with order parameters of dimension one
as well as dimension three.  Introductory references on supersymmetry
dynamics include Seiberg's lectures in this volume, and \cite{peskintasi,dinebook,dinemasonreview}.

There is a huge literature on gaugino condensation, but one should ask:  what is the essence of this
phenomenon.   Almost all discussion focuses on the fact that an
$SU(N)$ gauge theory without matter has a $Z_N$ discrete symmetry, broken by
gaugino condensation, a non-zero value of the dimension three order parameter,
\beq
\langle \lambda \lambda \rangle = \Lambda^3 e^{2 \pi i k \over N}.
\eeq
More generally, any non-abelian gauge theory without matter exhibits such a phenomenon.

But, if we forget the details of the models, we might extract three features:
\begin{enumerate}
\item  Breaking of a discrete $R$ symmetry.
\item  All scales arise through dimensional transmutation
\item  Order parameter of dimension $3$.
\end{enumerate}

If we relax the second item, then supersymmetric QCD with massive quarks already breaks a discrete $R$ symmetry, and possesses
gauge-invariant order parameters of dimension two.  But for thinking about supersymmetric models and dynamical supersymmetry breaking,
it is more interesting to relax the third item, i.e. we define gaugino condensation as\cite{dk}:
\begin{enumerate}
\item  Breaking of a discrete $R$ symmetry.
\item  All scales arise through dimensional transmutation
\end{enumerate}

A simple class of generalizations with gauge invariant
order parameters of dimension one is provided by
supersymmetric QCD with $N$ colors and $N_f$ flavors, $N_f < N$, and with
$N_f^2$ gauge singlet chiral fields, $S_{f,f^\prime}$.  For the superpotential, take:
\beq
W = y S_{f f^\prime} \bar Q_f Q_f^\prime +\lambda {\rm Tr} S^3.
\label{ssuperpotential}
\eeq
To simplify the writing, we have assumed an $SU(N_f)$ flavor symmetry; this is not necessary to any of our considerations
here.  This theory
possesses a $Z_{2(3N-N_f)}$ $R$ symmetry.  This can be seen by noting that an instanton produces $2N$ gaugino
 zero modes, and $2 N_F$ fermionic ($Q$ and $\bar Q$) zero modes.
 This symmetry is spontaneously broken by $\langle S \rangle;~ \langle \bar Q Q \rangle; ~\langle W_\alpha^2 \rangle$, $\langle W \rangle$.

The dynamics responsible for this breaking can be understood along the lines developed in Seiberg's lectures.  Suppose, for example, that
$\lambda \ll y$.  Then we might guess that $S$ will acquire a large vev, giving large masses to the quarks,
\beq
m_q = y s.
\eeq
In this case, one can integrate
out the quarks, leaving a pure $SU(N)$ gauge theory, and the singlets $S_{f,f^\prime}$.  The singlet superpotential follows by noting that the scale, $\Lambda$,
of the low energy gauge theory depends on the masses of the quarks, which in turn depend on $S$.  So
\beq
W(S) = \lambda S^3 + \langle \lambda \lambda \rangle_S.
\eeq
\beq
\langle \lambda \lambda \rangle = \mu^3 e^{-3{8 \pi^2 \over b_{LE} g^2(\mu)}}
\eeq
$$~~~~= \mu^3 e^{-3{8 \pi^2 \over g_{LE} g^2(M)} + 3{b_0 \over b_{LE}} \ln(\mu/M)},$$
where $\mu$ is the scale at which we match the couplings of the high and low energy theories, $\mu = m_q$, and
\beq
b_0 = 3N - N_F;~b_{LE} = 3N.
\eeq
So
\beq
\langle \lambda \lambda \rangle = M^{3N-N_f \over N} e^{-{8 \pi^2 \over N g^2(M)}} \mu^{N_f \over N}.
\eeq
In our case, $\mu = yS$, so the effective superpotential has the form
\beq
W(S) = \lambda S^3 + (yS)^{N_f/N} \Lambda^{3-N_f/N}.
\eeq
This has roots
\beq
S = \Lambda \left ({y^{N_f/N} \over \lambda} \right )^{N \over 3N - N_F}
\eeq
times a $Z_{3N-N_F}$ phase.

Note that this analysis is self-consistent; $S$ is indeed large for small $\lambda$.
The dynamics in other ranges of couplings has alternative descriptions, but the result that the
discrete symmetry is spontaneously broken, while supersymmetry is unbroken, always holds.

\subsubsection{Gauge Mediation/Retrofitting}

Given our models of gaugino condensation, it is a simple matter to generate the various dimensionful couplings of O'Raifeartaigh models dynamically.
In the model of \ref{simplestor}, for example, we can make the replacements:
 \beq
X (A^2 - \mu^2) + m AY
\label{retrofittedor}
\eeq
$$~~~~\rightarrow{X W_\alpha^2 \over M_p}+ \gamma S A Y.
 $$
Note that $\langle W \rangle  \approx  \Lambda^3$, $\langle S \rangle \sim \Lambda$, and
$m^2 \gg f$.
SUSY breaking is metastable, as in our earlier perturbed O'Raifeartaigh models (again, the supersymmetric vacua are far away).

\vskip .2cm
\noindent
{\bf Exercise:}   Verify that eqns. \ref{retrofittedor},\ref{ssuperpotential} respect a suitable discrete $R$ symmetry.
\vskip .2cm

\subsubsection{Gauge Mediation and the Cosmological Constant}

A traditional objection to gauge mediated models\footnote{I first heard this from Tom Banks.}
is that the smallness of the c.c. requires a large constant in $W$, unrelated to anything else.

But we have just seen that in retrofitted models, one naturally expects $\langle W \rangle \approx F M_p^2$, i.e. of the correct order of magnitude to (almost) cancel
the susy-breaking contributions to the c.c.  This makes retrofitting, or something like it,
almost inevitable in gauge mediation.

Other small mass parameters, such as the $\mu$-term, arise readily from dynamical breaking of the discrete $R$ symmetry.  For example
\beq
W_{\mu} =
{S^2 \over M_p} H_U H_D\eeq
gives rise to a $\mu$ term due to the expectation value of $S$.

In traditional approaches to gauge mediation, the $\mu$ term is problematic, not so much because it is hard to generate the term itself, but because $B_\mu$
tends to be too large.  If, for example, $\mu$, like the soft breaking masses is generated at, say, two loop order, $B_\mu$ is also typically generated
at two loop order, meaning that $B_\mu \gg \mu^2$.  This tends to lead to problematic hierarchies of mass scales.
But in the present case,
because $S$ has only a tiny $F$-component, the
corresponding $B$ term is extremely small.  A larger contribution arises from renormalization group running from the messenger
scale to the TeV scale.  This is suppressed by a loop factor but enhanced
by a logarithm.  As a result, the expectation value of $H_D$ is suppressed relative to that of $H_U$, leading to a prediction of
a large value of $\tan \beta$,
\beq
\tan(\beta) = {\langle H_U \rangle \over \langle H_D \rangle}.
\eeq
With these ingredients, it is almost {\it too} easy
to build realistic models of gauge mediation/dynamical supersymmetry breaking with all scales dynamical, no $\mu$ problem,
and prediction of a large $\tan \beta$.

\subsubsection{R Symmetry Breaking in Supergravity}

Even in supergravity theories, the scale of the superpotential is small compared to the Planck scale.  As we have already
mentioned, in the KKLT construction this is not particularly natural; one has to assume a selection
of small $W_0$.  One might hope to account
for this phenomenon through $R$ symmetries.

In supergravity (superstring) theories, there are natural candidates for Goldstino fields.  These are the fermionic partners
of the (pseudo) moduli.  Classically, by definition, these fields have vanishing
superpotential.     They might
acquire a superpotential through non-perturbative effects:
\beq
W = f~ M_p~ g(X/M_p).
\eeq
For
$X \ll M_p$, there might be an approximate $R$, along the lines required by Nelson and Seiberg,
perhaps due to discrete $R$ symmetries.
But it is unclear how one can get a {\it large} enough $W$ under these circumstances to cancel the c.c.;
$\langle W \rangle$ would be suppressed by both $R$ breaking and susy breaking.
Alternatively, one could, again, retrofit scales, as in the low scale models.

These sorts of questions motivate study of $W$ itself as an order parameter for $R$-symmetry breaking.
In the next section, we will prove a theorem about the size of $\langle W \rangle$ in models with
continuous $R$ symmetries.

\subsubsection{A Theorem About the Superpotential}

While not critical to our subsequent discussion, it is interesting that there is a quite general
statement that one can make about $\langle W \rangle$ in any globally supersymmetric
theory with a continuous $R$ symmetry.

\noindent  {\bf  Theorem}\cite{dfk}:
In any theory with spontaneous breaking of a continuous R-symmetry and SUSY:
$$
|\langle W\rangle|\leq {1 \over 2} |F|f_a
$$
where $F$ is the Goldstino decay constant and $f_a$ is the R-axion decay constant.

We will content ourselves with demonstrating the result for O'Raifeartaigh models with quadratic Kahler
potentials and arbitrary superpotentials; the bound can be shown to hold
quite generally, even in strongly coupled theories.

Consider a generic renormalizable O'Raifeartaigh model with an R-symmetry $\Phi_i\rightarrow e^{i q_i \xi}\Phi_i$.
$$
K=\sum_i \Phi_i \bar{\Phi}_i,\;\;\;W(\Phi_i)=f_i \Phi_i +m_{i j}\Phi_i \Phi_j+\lambda_{i j k}\Phi_i \Phi_j \Phi_k
$$
If the theory breaks SUSY at $\phi_i^{(0)}$ then classically it has a pseudomoduli space parameterized by the goldstino superpartner\cite{ray,ks}.
$$
G=\sum_i \left({\partial W\over \partial\phi_i}\right) \psi_i,\qquad~~~
\phi=\sum_i \left({\partial W\over \partial\phi_i}\right) \delta \phi_i,\qquad
$$

\vskip .3cm
\noindent
{\bf Exercise:}  Verify that $\phi$ is massless.

\vskip .3cm

Wherever the R-symmetry is broken there is also a flat direction corresponding to the R-axion.
Define two complex vectors $w_i= q_i \phi_i$ and $v_i^{\dagger}= {\partial
W\over\partial \phi_i}$.
Since
the superpotential has R-charge 2,
$$2\langle W\rangle=\sum_j q_j \phi_j{\partial W(\phi_i)\over
\partial \phi_j}=\langle v, w\rangle.$$
On the pseudomoduli space we can write
$$
|F|^2=\sum_i \left({\partial W\over
\partial \phi_i}\right)\left({\partial W\over
\partial \phi_i}\right)^* = \langle v,v\rangle.
$$
Parameterizing $\phi_i(x)=\langle\phi_i(x)\rangle e^{i q_i a(x)}$ we obtain for the R-axion kinetic term:
$$
\left(\sum_i|\phi_i(x)|^2 q_i^2\right) (\partial a)^2 \Rightarrow f_a^2=\langle w,w\rangle
$$
Then by the Cauchy-Schwarz inequality:
$$
4|\langle W\rangle|^2=|\langle v,w\rangle|^2\leq \langle v,v \rangle \langle w,w \rangle=|F|^2f_a^2
$$
which is the bound to be established.

It is worth noting that:
\begin{itemize}
\item
The bound is saturated if $v\propto w$, in which case the $R$-axion is the Goldstino superpartner.
\item  Adding gauge interactions strengthens the bound because the $D$ terms contribution to the potential makes $|F|^2$ larger.
\end{itemize}
The proof can readily be generalized to arbitrary Kahler potential.  Further input is needed to prove the result in full generality, i.e. for strongly coupled
field theories as well.

\section{Lecture 3.  Supersymmetry in String Theory}

At this school, and in most string papers, it is taken as a given that low energy supersymmetry is a consequence of string theory.
But, as we will see, this is by no means self-evident.  In this lecture, we will outline some of the issues.  We will argue that if one
could assert that low energy supersymmetry {\it is} an outcome of string theory, this is a dramatic prediction.  Even more exciting
would be to make some statement about the form of supersymmetry breaking.  But it is quite possible that string theory
predicts no such thing.

In this lecture we will consider:
\begin{enumerate}
\item  What might it mean for string theory to make contact with nature.   We will argue that the {\it landscape}
is the only plausible setting we have contemplated to date.
\item  The elephant in the room:  The cosmological constant.
\item  The Banks/Weinberg proposal\cite{bankscc,weinbergcc}.
\item  The Bousso-Polchinski model\cite{bp}  (string theory fluxes) as an implementation (details for Denef).
\item  KKLT as a model.  What serves as small parameter?  Why is a small parameter important?
\item  Distributions of theories.
\item  Supersymmetry in string theory and the landscape
\item  KKLT as a realization of intermediate scale susy breaking.
\item  A new look at susy breaking in the KKLT framework.
\item  The landscape perspective on intermediate scale breaking.
\item  Assessment
\item  Discrete symmetries in string theory and the landscape
\item  Strong CP and axions in string theory and the landscape (KKLT)
\item  Assessment:  string theory predictions(?)
\end{enumerate}

\subsection{String Theory and Nature}

In Miriam Cvetic's lectures, you heard how string theory can come close to reproducing many features of the Standard Model:  the gauge group,
the number of generations, and at least some features of Yukawa couplings.  The constructions she described typically come with less desirable features, especially extra
massless particles.

But these constructions raise an obvious question:  there seem to be myriad possibilities.  What principle governs
which nature chooses, if any?  Why, say, intersecting branes, and not heterotic constructions, non-geometric models, F theory constructions, or something
perhaps not yet known.  What sets a particular value of the allowed fluxes?
One possible response is that eventually we will find {\it the} model which describes everything, work out its consequences, and make other predictions.
Another is that we might find some {\it principle} which provides the answer, pointing to a unique string vacuum state.  Finally, there is the point
of view advocated by Banks at this school, that the different string ``vacua" are actually different theories of quantum gravity; indeed, there are simply many different theories
of quantum gravity, just as there are
many possible different field theories.

But there is at least one fact which points to a different possibility; this is is the cosmological constant.

\subsection{The cosmological constant problem}

At one level the cosmological constant problem is simply one of dimensional analysis:  one would naturally expect that the c.c. would be of
order some microscopic scale to the fourth power.  $M_p^4$ would give a result $120$ orders of magnitude larger than the observed cosmological
constant, but even $M_Z^4$ would miss the observed value by more than $55$ orders of magnitude.
Were we to suppose that, for some reason, the cosmological constant is zero classically, quantum corrections
would seem, inevitably, to be huge.
The problem is illustrated by our earlier expression for the vacuum energy, \ref{cwpotential}, which is quartically divergent.  Typically, in string models, if
there is no supersymmetry, one obtains
a result of the predicted order of magnitude, with a suitable cutoff (e.g. the string scale; such a calculation was
first carried out by Rohm\cite{rohm}). (In practice, since there are typically moduli, and certainly
moduli in any case where weak coupling computations make sense, one is actually calculating a potential for the moduli.)
So there is no evidence that string theory performs some magic with regards to this problem.

\subsubsection{Banks, Weinberg:  A proposal}

Now I will use words that Tom told you one should never use, they are at the heart of a brilliant idea which he put forward\cite{bankscc} and Steven Weinberg turned into
a dramatic prediction\cite{weinbergcc}.  In order to understand the smallness of $\Lambda$,
suppose that the underlying theory has many, many vacuum states, with a more or less uniform
distribution of c.c.'s (Bousso and Polchinski dubbed such a distribution a ``discretuum"\cite{bp}).  Suppose that the system makes transitions between these states, or in some other
way samples all of the different states (e.g. they all exist more or less simultaneously).  Now imagine a star trek type figure, traveling around this vast universe.
(This requires all sorts of superluminal phenomenon, but we won't worry about this, which is to say that we don't understand
the issues too well, not that they are unimportant!).  Most universes will not be habitable; the c.c. will be of order, say,
the Planck scale.  But sometimes the c.c. will be smaller, and the universe will be comparatively flat.  Under what conditions might this
star trek character find intelligent life?  That's a tough question.  As a proxy, Weinberg asked:   under what circumstances, assuming all of the other
laws of nature are the same, will this observer find galaxies.  Weinberg noted, first, negative c.c. is unacceptable; if the resulting Hubble parameter is not much smaller than
(one over) a billion  years or so,
the universe will undergo a big crunch long before galaxies, much less life, form.  Similarly, for large, positive c.c., the would be structures will be ripped apart before galaxies
can form.  This sort of argument gives a c.c. about $100$ times as large as observed.  . (This was actually a prediction).  This is remarkably good (Weinberg was originally rather
negative about the result).  On a log scale, it's excellent.  More refined versions of the argument\cite{refinedcc} reduce this discrepancy by an order of magnitude.

\subsection{Bousso-Polchinski and KKLT}

Bousso and Polchinski (BP) proposed that such a discretuum might arise in string theory as a result of fluxes.  The idea is simply that if there are many possible fluxes (say $\chi$), each
of which can take $N$ different values, one has of order $N^{\chi}$ states.  If, say, $\chi \sim 100$ and $N \sim 10$, this number is huge.
 As a model,  BP assumed that there is one (meta-)stable {\it vacuum} or
{\it state} for each flux choice, but it is not clear this is reasonable.  It is certainly not clear, for example,
 that moduli are fixed for every choice of fluxes.  KKLT put forward a plausible implementation of the BP proposal in string theory which included a dynamical mechanism
 to fix all of the moduli, and a candidate
for a small parameter which would allow exploration of some states.  Central to their scenario are the 3-form fluxes.  Again, if there are  $\chi$ types of fluxes, each
taking $L$ values, of order $L^{\chi}$ states.  For known Calabi-Yau manifolds, this number
can easily be enormous, large enough to give the sort of discretuum required.

The elements of the KKLT proposal are readily enumerated; more detailed aspects of the construction are described in Denef's lectures at this school.
\begin{enumerate}
\item  The first ingredient is a IIB Orientifold of CY, or F-theory on CY Four-Fold.
with $h_{2,1}$ complex structure moduli, $h_{1,1}$ Kahler moduli and the dilaton (and their superpartners)
\item  Three form fluxes fix the complex structure moduli and dilaton, along lines discussed in \cite{beckerbecker, otherfluxes,gkp}.
\item  The fixing of the complex structure moduli leaves a low energy theory which is approximately supersymmetric,
with a small parameter, $W_0 = \langle W \rangle$.
\item  Non-perturbative dynamics fix the remaining Kahler moduli in terms of the small number, $W_0$.
\item  Additional branes are required in order to obtain matter, and perhaps for susy breaking.
\end{enumerate}

In the spirit of these lectures, we will focus on the low energy effective theory for the Kahler moduli and other light fields.  The details of steps 1-2 are discussed in
Denef's lectures.

\subsubsection{Fixing the Kahler moduli}

For simplicity, we'll suppose that there is only one Kahler modulus (new issues which arise in the presence of multiple Kahler moduli are discussed
in \cite{acharya, raby,dinemultiaxions}; these will be discussed in section \ref{axions}).  The Kahler modulus sits in a chiral multiplet, $\rho$; the other scalar in the multiplet is an axion, which, from the microscopic
point of view, respects a discrete shift symmetry.  In other words,
\beq
\rho = \rho + i a
\eeq
and the theory is symmetric under $a \rightarrow a + 2 \pi$ (in a suitable normalization).
The low energy theory, after integrating out the complex structure moduli
is described by a {\it supersymmetric} effective action with
\beq
W = W_0 ~~~~~K = -3 \ln(\rho + \rho^\dagger)
\eeq
as well as some additional fields (gauge fields and charged matter) localized on branes.  As it stands, this low energy theory breaks supersymmetry,
with a potential which, at the classical level, is independent of $\rho$ (a so-called ``no-scale" model).

In general, the parameter $W_0$ is of order one.  But KKLT and Douglas\cite{douglasdenef1} argued that
the large number of possible flux choices would give rise to a distribution of the parameters of this low energy lagrangian.
$W_0$, in particular, would be distributed uniformly as a complex variable.  If there
are many states, then, in some, $W_0$ will be small.  $W_0$
serves as the small parameter of the KKLT scenario (note that if the number of states is finite, $W_0$ cannot be arbitrarily small, so there is not really a
systematic expansion).

In order to account for the stabilization of $\rho$,
it is assumed that either stringy instanton effects or effects in the low energy theory, such as gaugino condensation, give rise to an additional term
in the superpotential
\beq
W = W_0 - A e^{-\rho/b}.
\eeq
 For $W_0$ small, the superpotential has a supersymmetric stationary point:
\beq
D_\rho W \approx aA/b e^{-\rho b} -  {3\over \rho + \rho^\dagger}W_0 = 0
\eeq
\beq
\rho = \rho_0  \approx -b \ln(\vert W_0 \vert).
\label{rhosolution}
\eeq
This nominally justifies a large $\rho$ expansion, i.e. the $\alpha^\prime$ expansion (ref. ~\cite{gkp} proposed a  mechanism to achieve
weak string coupling).





\subsubsection{Supersymmetry Breaking in KKLT}

So far in this discussion we have not accounted for supersymmetry breaking.
KKLT proposed that $\overline{D3}$ branes could break susy.   Indeed, such branes, if present, would appear
to break supersymmetry {\it explicitly} in the low energy theory.  This is somewhat confusing,
since it is not clear how this could be described in an effective low energy theory.  After all,
one is supposing that a system with a light gravitino is not supersymmetric, which would appear to be inconsistent.
The resolution may, perhaps, be that for such (warped) branes, the low energy theory is not a four dimensional
theory with a finite number of fields.  One can avoid this issue by simply hypothesizing there is another (four dimensional) field
theory sector, again perhaps localized
on a brane, which
spontaneously breaks susy, perhaps along lines of models we discussed in the previous lecture.  What is important is that this extra sector gives
a positive contribution to the cosmological constant (you can fill in Banks objections at this point).

So KKLT provided a plausible (but hardly rigorous) scenario to understand:
\begin{enumerate}
\item  The existence of a large number of (metastable) states.
\item  Fixing of moduli.
\item  Breaking of supersymmetry.
\item  Distribution of parameters of low energy physics, including the c.c. and the scale of supersymmetry breaking.
\end{enumerate}

Based on this analysis, the likely existence of a ``landscape" of vacua in string theory has become widely (though certainly
not universally) accepted.
Adopting this viewpoint, Douglas and Denef embarked on a program of studying the statistics of KKLT-like
vacua.  The most primitive question they investigated was counting, but they also studied distributions of various
quantities in addition to $W_0$, such as the supersymmetry-breaking scale\cite{douglasdenef1,douglasdenef2}.

If we adopt this view, we still would like to know which features of the KKLT construction might be generic.
Are typical states in the landscape approximately supersymmetric?  Or is supersymmetry, perhaps, exponentially rare?
Within the supersymmetric states, are features of the spectrum similar to those of KKLT?  For example, the modulus,
$\rho$, is parameterically heavy (by a power of $\rho$) compared to $m_{3/2}$.  Is this typical?  Are moduli
often heavy in this sense?

\subsection{What might we extract from the landscape?}

If the landscape viewpoint is correct, we
can't hope to find ``the state" which describes
the world around us.  I like to describe this problem as follows.  Imagine an army of graduate students.  Each is given
a state to study, specified by a bar code.  Graduate student A calculates the c.c. to fourth order and finds it's very small.
She becomes exited.  She goes on to third order, which takes her five years.  She is even more excited, as she is even
closer to the observed value.  She keeps going.  After 40 years she completes eighth order.  But, oh well, she is not
within errors.  She returns to her adviser, and is given another bar code.  Wati Taylor has described, at this school,
a more plausible program involving searches for correlations among various quantities, such as numbers of generations and features of Yukawa couplings.

But I would suggest that the most promising questions are those
connected with questions of naturalness.  This is precisely because the
phenomena we are trying to explain seem at first sight unlikely.
These are
\begin{enumerate}
\item  The c.c. (already discussed)
\item  The strong CP problem
\item  Fine tuning of the electroweak scale.
\item  Cosmological issues such as inflation.
\end{enumerate}
There has been much work on the last of these, but it is very unclear what might be generic.  I'll focus on the second and third.
The reader should be warned that we are entering, here, a zone of (possibly biased) speculation.

\subsection{Supersymmetry in the Landscape}

At first sight, supersymmetry would seem special.  Even if it is easier to explain hierarchies, we are talking about such large numbers of states that
the number of non-susy states exhibiting huge hierarchies, could well overwhelm those in which the scale arises naturally\cite{douglas,susskind}.
To consider this question, we can
divide the landscape into three branches:
\begin{enumerate}
\item  States with no supersymmetry
\item  States with approximate supersymmetry
\item  States with approximate supersymmetry and discrete $R$ symmetries.
\end{enumerate}

One possible argument against the first branch and in favor of branches two and three invokes stability (now I will invoke a simple-minded version
of Banks' concerns about ``states").  It is usually said that tunneling amplitudes are naturally small, but this requires the existence of some small parameter
(a small coupling, a small ratio of energy splitting to barrier height, a small ratio of energy scales to the distance over which one tunnels...).
In the landscape, a typical state with small cc would be surrounded by vast number of states with large, negative cc.   What prevents decays to big crunch
space-times?  In a typical
state, one expects no parameter which would account for the smallness of decay rates.  Note that {\it every} decay channel
must be suppressed
and that there are potentially an exponentially large number of
such channels\cite{dinestability} .  One can imagine various features which might be typical
of many landscape states and which might suppress decays:
\begin{enumerate}
\item  Weak string coupling
\item  Large compactification volume
\item  Warping
\item  Supersymmetry
\end{enumerate}
Items 1 and 3 appear not to be lead to metastability in generic situations\cite{dinestability}.  Large compactification volume, to be effective,
requires that the volume scale as a power of the typical flux.  No non-supersymmetric landscape model
studied to date has this feature, but perhaps it is possible.  Supersymmetry {\it does} generically lead to stability.  It is, indeed, a theorem that, with unbroken supersymmetry,
Minkowski space is stable.  With small supersymmetry breaking, one finds that
the decay amplitudes vanish, or are exponentially small\cite{susytunneling}:
\beq
\Gamma \approx e^{-M_p^2/m_{3/2}^2}.
\label{susydecay}
\eeq

So, even though the conditions for a supersymmetric vacuum (e.g. conditions for a supersymmetric
stationary state of some complicated action) may be special, requiring some degree of metastability -- that, say,
some stationary point of the effective action be in any sense a ``state" -- might favor supersymmetry.
One can study this question in toy landscapes; whether one make a definite statement requires
a deeper understanding of the landscape.

Even if one accepts that some degree of supersymmetry is generally needed to account for stability, by itself this would not explain
why very low scale supersymmetry should emerge.  After all, examining equation \ref{susydecay}, it is clear that if $m_{3/2} /M_p \sim 10^{-3}$,
the decay amplitude is already unimaginably small.  In terms of our classification of states, in other words, stability perhaps accounts for why
we are not on the first (non-supersymmetric) branch, but not why {\it very} low energy supersymmetry should be favored.
But from facts which are understood about supersymmetric landscapes, it would appear that the explanation
could simply be conventional naturalness.  For example, studying IIB landscapes, Douglas and Denef found  instances with a uniform distribution
of gauge couplings\cite{douglasdenef1}.  It would seem plausible that, in a significant fraction of states,
supersymmetry is broken dynamically, i.e.
\beq
m_{3/2} \propto e^{-{8 \pi^2 \over b g^2}},
\eeq
The uniform distribution of $g^2$ then implies a logarithmic distribution of $m_{3/2}^2$ (as in the original arguments for hierarchies, e.g. \cite{susskindtechnicolor}).
Coupling this with the assumption of a uniform distribution of $W_0$, and requiring small c.c., gives a distribution of supersymmetry breaking
scales which is uniform on a log scale.  If $W_0$ is itself fixed dynamically, as might be the case in models with discrete $R$ symmetries,
then low scale supersymmetry breaking is favored even more strongly\cite{dgt}.

There are further questions one can sensibly ask, even given our limited understanding of these issues.  For example, one might argue that the third branch,
with $R$ symmetries, while favoring low energy supersymmetry breaking, is itself disfavored.  After all, in order that the low energy theory exhibit an $R$ symmetry,
in the case of flux vacua, it is necessary to set to zero all fluxes which transform non-trivially under the symmetry.  Given the assumption that the vast number of states
arises because there are many types of fluxes, which take many values, this has the effect, typically, of reducing the number of flux types by an order one fraction\cite{dinesun},
and thus reducing the number of available states by an exponential amount (say $e^{600} \rightarrow e^{200}$).  On the other hand,
one can advance rather primitive cosmological arguments that such states might be favored\cite{dinesymmetries}.

\subsubsection{Axions in the Landscape}
\label{axions}

The landscape suggests the possibility that all quantities relevant to low energy physics are randomly distributed, unless selected
anthropically.  But not all of the quantities in our low energy effective
theory look random.  This applies to the quark and lepton mass matrices, which exhibit curious patterns.  Most dramatically, though, it applies to the $\theta$ parameter
of QCD, a pure number, which, if random, one might expect to be of order one.  Instead, we know $\theta < 10^{-9}$.  No known anthropic argument
would seem to select for such a small value, which would seem otherwise highly improbable\cite{bdg,donoghue}.

A natural response is that in string theory, axions seem ubiquitous, so perhaps there is an automatic solution here.  But in the case of KKLT, we have just seen
that, while there is an axion candidate, it is fixed at the same time that $\rho$ is fixed (it is fixed by the phase of $W_0$ in eqn. \ref{rhosolution}).  As a result, it is very
massive, and can play no role in solving the strong CP problem.  This has lead to significant pessimism about the strong CP problem in string theory.

But perhaps KKLT is a little too simple as a model.  In particular, typical string
compactifications have multiple Kahler moduli.  This leads to more interesting possibilities\cite{acharya,raby,dinemultiaxions}.
Consider the case of two Kahler moduli, $\rho_1,\rho_2$.  It is now possible that one linear combination of these is fixed in an approximately
supersymmetric fashion, as in KKLT, while the other is fixed by supersymmetry-violating dynamics.  The ``axionic" component of this multiplet
need not be fixed by these dynamics.  The requirements can be understood by examining a simple model:
\beq
W = W_0 + A e^{-{(n_1 \rho_1 + n_2 \rho_2 + n_3 \rho_3) \over b}} + C e^{- \rho_1} + D e^{-\rho_2} + \dots
\eeq
The second term might be generated by gaugino condensation, say, on some brane, with $b > 1$, while the third and fourth could be generated by
high scale, ``stringy" instantons.  The first term would fix, in a supersymmetric fashion one linear combination
of moduli.  In other words, the real and imaginary parts  of one field would gain mass approximately
supersymmetrically.  Supersymmetry breaking effects could fix the real part of the other linear combination, leaving a light axion, which would obtain
mass only from QCD and the (exponentially suppressed) third and fourth terms in $W$.

More precisely, in this model one has a ``heavy", complex field,
\beq
\Phi = n_1 \rho_1 + n_2 \rho_2 + n_3 \rho_3
\eeq
and two light fields.
As in the KKLT model,
\beq
\Phi \sim b \log(\vert W_0 \vert).
\eeq
The $\Phi$ field has mass of order $\Phi m_{3/2}^2$.  At lower
energies, one has one light modulus, $\phi$ (as well as any additional matter fields on branes, etc.); the effective
action consists simply of the Kahler potential for these fields and a constant superpotential.  Supersymmetry
breaking in this theory (which can be due to $\phi$ itself) can fix the real part of $\phi$, leaving
a light axion.

Apart from possibly accounting for the QCD axion, the
fact that there might be many $\rho_i$-type fields  could give rise to the ``axiverse" of ref. \cite{axiverse}.  The idea is that if the masses of axions
arise from $e^{-\rho_i}$ type factors, their masses might be roughly uniformly distributed in energy scale, leading
to interesting possible cosmological and astrophysical phenomena.

\subsection{A Top Down View of TeV Physics}

This combination of considerations provides a coherent, principled   (though not necessarily true!)
picture in which perhaps low energy supersymmetry is about to be found at LHC.  But it could well be a house of cards.  It stands
on many shaky assumptions, most crucially the existence of a landscape, and secondarily a landscape which resembles the limited
sets of configurations which have been studied by string theorists.
Perhaps even within a landscape framework
some other phenomena (warping, large extra dimensions, technicolor?) is responsible for electroweak symmetry breaking.  Or worse, the landscape idea
points to the alternative possibility that the electroweak scale is determined anthropically and
we are about to find a single light Higgs.  Arguments have been offered that this may not be the case\cite{antianthropichiggs},
but the stress should be on {\it may}.  It is one thing to dislike anthropic arguments; another
to rule them out.  I hope I have outlined some questions and some possible approaches, but perhaps your young, fresh minds
will come up with better ways of thinking about these questions.  And perhaps, within a few years (maybe even only one!), we will have experimental verification of some of these
ideas, or unexpected clues as to what physics lies Beyond the Standard Model.

\bibliographystyle{ws-rv-van}
\bibliography{tasi_2010_arxuv.bbl}


\end{document}